\documentclass{article}
 
\usepackage{graphicx}
\usepackage{booktabs}
\usepackage{amsmath}
\usepackage{comment}
\usepackage{fullpage}

\title{
A Study of the Provincial Road Networks of Canada as Complex Networks}
\author{Srivatsan Vasudevan \qquad Asish Mukhopadhyay \qquad Sudarshan Sundararajan}
\author{Srivatsan Vasudevan, Asish Mukhopadhyay and Sudarshan Sundararajan\\
	School of Computer Science \\
	Univesity of Windsor\\
	Ontario, Canada}

\date{}

\begin{document}
\maketitle

\begin{abstract}
	In this paper, we report on a study of the road networks of the provinces of Canada as complex networks. A number of statistical features have been analysed and compared with two random models. In addition, we have also studied the resilience of these road networks under random failures and targeted attacks. 
\end{abstract}

\section{Introduction}
Statistical properties of protein-protein interaction networks, social networks, the world-wide web etc., have been studied in the framework of complex networks, leading to
interesting insights into the function and properties of these networks. Motivated by this, we study the statistical properties of road networks of the 10 provinces of Canada. In addition, such a study would be an invaluable resource document for planning new transport infrastructure to address the needs of a dynamically growing population.\\    

 A random graph provides the mathematical model that serves as a benchmark or null hypothesis for studying real networks. Thus in this study, the statistics of each of the ten road networks is compared with a random graph, with the same number of nodes and edges, generated in Erdos-Renyi \textit{random graph model} and another generated in the more general and versatile random graph model, known as the \textit{configuration model}. \\
 
 In the next section, we describe the data collection process; in the following section, we study some select statistical features of the road networks. Motivated by potentially catastrophic disruptions (climatic, for example) of a road network, in section 4 we investigate the resilience of the road networks under random and targeted disruptions. \\

\section{Data Extraction Methodology}
\begin{figure}[!htbp]
    \centering
    \includegraphics[width=1\textwidth]{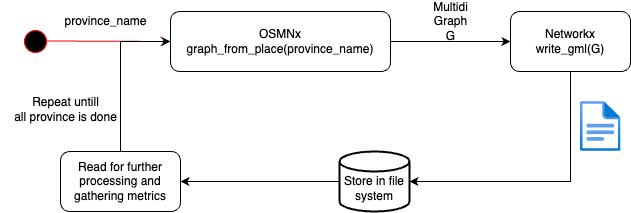}
    \caption{Flow Diagram of data extraction}
    \label{fig:data_extraction}
\end{figure}
The OSMnx \cite{boeing2017osmnx} Python module is a powerful toolkit for extracting, modeling, and analyzing geographical networks, with a special emphasis on road networks. OSMnx uses data from OpenStreetMap to allow users to choose a geographical location, such as a province, and returns the accompanying road network as a graph. Figure \ref{fig:data_extraction} shows the data flow diagram (DFD) of the process that extracts a geographical network using the OSMnx library. The initial input is a province name to the function {\tt graph\_from\_place(province\_name)}, which outputs a directed multi-graph $G(V, E)$. The nodes of $G$ represent the road intersections and the edges of $G$ represent the road segments that define this intersection. That $G$ is a directed multigraph means all the edges have directions and more than one edge can connect two arbitrary nodes $u, v$  of $G$. Next, using the versatile Networkx \cite{hagberg2020networkx} python package used in the analysis of networks, graph $G$ is saved into the file system as a GML (Graph Modelling Language) file. GML is a text-based format commonly used for representing directed or undirected graphs, making it suitable for storing complex network structures like a province’s road network. This makes the task of analyzing the networks easier and also provides a way to replicate the results discussed in the subsequent sections.\\

\section{Some Select Statistical Features}
In any network analysis, the first metric that merits scrutiny is the number of nodes and edges in the network. Table \ref{table:network-stats} encapsulates this information about the road networks of all 10 provinces. We note that there are big differences in the number of road segments and the number of intersections among the provinces. Ontario, Quebec, Alberta, British Columbia and Saskatchewan have more than 100k nodes or intersections whereas the other provinces have fewer, with Prince Edward Island having the least of them all with just over 10k intersections. The number of edges or road segments also follows a similar pattern. \\

\begin{table}[!htbp]
    \centering
    \begin{tabular}{l|c|c} 
        \toprule
        \textbf{Province} & \textbf{Number of Nodes} & \textbf{Number of Edges} \\
        \midrule
        Nova Scotia & 46,182 & 111,310 \\
        Ontario & 359,535 & 938,322 \\
        Manitoba & 70,694 & 197,154 \\
        Quebec & 265,932 & 682,993 \\
        Alberta & 256,831 & 631,772 \\
        Saskatchewan & 126,405 & 348,263 \\
        New Brunswick & 44,220 & 106,184 \\
        British Columbia & 133,872 & 321,925 \\
        Prince Edward Island & 10,078 & 25,172 \\
        Newfoundland and Labrador & 35,329 & 80,704 \\
        \bottomrule
    \end{tabular}
    \caption{Sizes of the Province Road Networks}
    \label{table:network-stats}
\end{table}
This is expected as the road network is built to cater to the population in these provinces and the population Table \ref{table:network-stats-population} below validates this correlation. The population data of each province has been obtained from the Statscan website \cite{statisticscanada}.
\begin{table}[!htbp]
    \centering
    \begin{tabular}{l|c|c|c} 
        \toprule
        \textbf{Province} & \textbf{Number of Nodes} & \textbf{Number of Edges} & \textbf{Population} \\
        \midrule
        Ontario&359535&938322&15911285\\
        Quebec&265932&682993&8984918\\
        British Columbia&133872&321925&5431355\\
        Alberta&256831&631772&4800768\\
        Saskatchewan&126405&348263&1225493\\
        Manitoba&70694&197154&1474439\\
        Nova Scotia&46182&111310&1069364\\
        New Brunswick&44220&106184&846190\\
        Newfoundland and Labrador&35329&80704&540552\\
        Prince Edward Island&10078&25172&176162\\
        \bottomrule
    \end{tabular}
    \caption{Network Size versus Population Size}
    \label{table:network-stats-population}
\end{table}

\newpage
\subsection{Clustering Coefficient}
The local clustering coefficient of a node $i$ quantifies the cliquishness (or connectivity) of a node's neighbourhood \cite{fagiolo2007}.

\[
c_u = \frac{T(u)}{2(\deg^{\text{tot}}(u)(\deg^{\text{tot}}(u) - 1) - 2\deg^{\leftrightarrow}(u))}
\]
where:
\begin{itemize}
	\item $T(u)$ is the number of directed triangles through node $u$.
	\item $\deg^{\text{tot}}(u)$ is the sum of the in-degree and out-degree of node $u$.
	\item $\deg^{\leftrightarrow}(u)$ is the reciprocal degree of node $u$.
\end{itemize}

A global estimate of this measure for the entire graph is obtained by computing the average of the local clustering coefficients of all the vertices in the graph as below:\\
\[
c = \frac{1}{N} \sum_{i \in V} c_i
\]
where $c_i$ represents local clustering coefficient of vertex $i$ and $N$ represents the total number of nodes in the graph. \\

From table \ref{tab:graph-clustering-coefficients} we observe that the average local clustering coefficient is high for each province in comparison with their respective random graph models. We note that the global clustering coefficient also shows a similar trend. This indicates the presence of clusters in all the provincial road networks. \\

\begin{figure}[h!]
	\centering
	\includegraphics[scale=0.7]{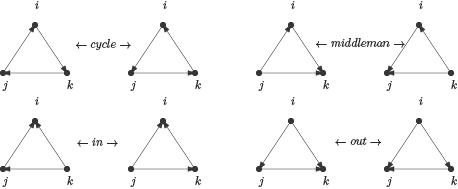}
	\caption{Directed triangles incident on vertex $i$}
	\label{fig:clusteringPatterns}
\end{figure}

In \cite{fagiolo2007}, eight different clustering coefficients are defined, depending on the directedness of each edge that makes up a directed triangle incident on vertex $i$. 
These in turn are grouped into four types: \textit{cycle, middleman, in and out}(see Fig.~\ref{fig:clusteringPatterns}). Since the road networks of the provinces are essentially undirected graphs, we have not computed these different types of clustering coefficients. 

\begin{table}[h!]
    \centering
    \begin{tabular}{l|c}
        \toprule
        \textbf{Graph Name} & \textbf{Global Clustering Coefficient} \\
        \midrule
        Nova Scotia & 0.037522069345074884\\
        Erdos Renyi Nova Scotia & 4.543921329682012e-05\\ 
        Configuration Model Nova Scotia & 3.3073079389826195e-05\\
        Ontario & 0.048251411459397396\\
        Erdos Renyi Ontario & 7.202935777344378e-06\\
        Configuration Model Ontario & 9.834130990624786e-06\\
        Manitoba & 0.045504921394801494\\
        Erdos Renyi Manitoba & 2.9501882205081896e-05\\
        Configuration Model Manitoba & 3.431960953108434e-05\\
        Quebec & 0.046020266257265526\\
        Erdos Renyi Quebec & 8.090116275667468e-06\\
        Configuration Model Quebec & 1.1003529008131684e-05\\
        Alberta & 0.036883182912724945\\
        Erdos Renyi Alberta & 8.561449353254079e-06\\
        Configuration Model Alberta & 9.896261224955968e-06\\
        Saskatchewan & 0.03776036534494564\\
        Erdos Renyi Saskatchewan & 1.8667895136389256e-05\\
        Configuration Model Saskatchewan & 2.178372238891995e-05\\
        New Brunswick & 0.04130921468990875\\
        Erdos Renyi New Brunswick & 5.171839900269941e-05\\
        Configuration Model New Brunswick & 5.255109732721673e-05\\
        British Columbia & 0.039608137638506\\
        Erdos Renyi British Columbia & 2.1391931574792806e-05\\
        Configuration Model British Columbia & 1.4299373388652718e-05\\
        Prince Edward Island & 0.040585359107387045\\
        Erdos Renyi Prince Edward Island & 0.0001865929158050611\\
        Configuration Model Prince Edward Island & 0.0003859184078473619\\
        Newfoundland and Labrador & 0.03286609049537554\\
        Erdos Renyi Newfoundland and Labrador & 4.4805584871819415e-05\\
        Configuration Model Newfoundland and Labrador & 7.652555771664716e-05\\
        \bottomrule
    \end{tabular}
    \caption{Global Clustering Coefficients}
    \label{tab:graph-clustering-coefficients}
\end{table}

\newpage
\subsection{Average Shortest Distance}
The next topological property studied is the average shortest distance that estimates the shortest distance between any two distinct nodes, picked at random. The distance measure here is the number of directed edges along the shortest path between any two nodes. This is calculated using the equation given below:
\[
D = \frac{1}{N(N-1)}\sum_{s,t \in V} d(s,t) 
\]
where $d(s,t)$ is a measure of the shortest path from $s$ to $t$, $V$ is the set of all nodes in the graph and $N = |V|$.\\

 If there is no path from a node $u$ to a node $v$, then the default shortest distance is set to a maximum integer value of 2147483647. From the table, we observe that, despite the size of the graph, the average shortest distance between vertices is small for all the provinces. Another intriguing observation is that for all the provinces the random graph constructed using the Erdos Renyi model has a significantly higher value, whereas the random graph generated using the configuration model has a value of the same order as the provinces. The Erdos Renyi model's high values can be attributed to the presence of many unreachable paths. \\

\begin{table}[!htbp]
	\begin{center}
    \begin{tabular}{lc} 
        \toprule
        \textbf{Graph Name} & \textbf{Avg Shortest Distance} \\
        \midrule
        Nova Scotia & 22.15222561 \\
        Erdos\_Renyi\_Nova Scotia & 10341.70208 \\
        Configuration\_Model\_Nova Scotia & 15.10254497 \\
        Ontario & 4.900450389 \\
        Erdos\_Renyi\_Ontario & 1073.49372 \\ 
        Configuration\_Model\_Ontario & 4.268804188 \\
        Manitoba & 44.67356109 \\
        Erdos\_Renyi\_Manitoba & 4492.89145 \\
        Configuration\_Model\_Manitoba & 34.36634661 \\
        Quebec & 5.435196569 \\
        Erdos\_Renyi\_Quebec & 1524.372326 \\
        Configuration\_Model\_Quebec & 4.281091444 \\
        Alberta & 16.01079995 \\
        Erdos\_Renyi\_Alberta & 1789.856991 \\
        Configuration\_Model\_Alberta & 13.86329372 \\
        Saskatchewan & 20.55355934 \\
        Erdos\_Renyi\_Saskatchewan & 2589.355722 \\
        Configuration\_Model\_Saskatchewan & 6.988198076 \\
        New Brunswick & 47.20932853 \\
        Erdos\_Renyi\_New Brunswick & 11069.10252 \\
        Configuration\_Model\_New Brunswick & 24.15828839 \\
        British Columbia & 15.93452592 \\
        Erdos\_Renyi\_British Columbia & 3651.772081 \\
        Configuration\_Model\_British Columbia & 13.77707738 \\
        Prince Edward Island & 295.8976174 \\
        Erdos\_Renyi\_Prince Edward Island & 44953.9689 \\
        Configuration\_Model\_Prince Edward Island & 147.9858415 \\
        Newfoundland and Labrador & 30.97069985 \\
        Erdos\_Renyi\_Newfoundland and Labrador & 15622.19806 \\
        Configuration\_Model\_Newfoundland and Labrador & 18.92512441 \\
        \bottomrule
    \end{tabular}
    \end{center}
    \caption{Average Shortest Distances}
    \label{table:shortest distance}
\end{table}

Average shortest distance values are ranked among provinces and are shown in the table \ref{table:ranked shortest distance}. We observe that provinces with greater numbers of nodes and edges have smaller average shortest distances. Prince Edward Island has the highest value. This pattern is expected because a higher number of nodes and edges are required to support the population and their needs.\\

\begin{table}[!htbp]
    \begin{tabular}{lc}
        \toprule
        \textbf{Province} & \textbf{Avg Shortest Distance} \\
        \midrule
        Ontario & 4.900450389 \\
        Quebec & 5.435196569 \\
        British Columbia & 15.93452592 \\
        Alberta & 16.01079995 \\
        Saskatchewan & 20.55355934 \\
        Nova Scotia & 22.15222561 \\
        Newfoundland and Labrador & 30.97069985 \\
        Manitoba & 44.67356109 \\
        New Brunswick & 47.20932853 \\
        Prince Edward Island & 295.8976174 \\
        \bottomrule
    \end{tabular}
    \centering
    \caption{Average Shortest Distances Ranked}
    \label{table:ranked shortest distance}
\end{table}

A small-world network has the following two distinctive characteristics (see\cite{watts1998collective}).

\begin{itemize}
    \item High clustering coefficient when compared to a random network of the same size.
    \item The average node-to-node distance is approximately $\log(N)$
\end{itemize}
    .
All the province networks satisfy the first criteria, but not all provinces have an average shortest distance in the range of $\log(N)$. Specifically, the province of Prince Edward Island has a high average node-to-node distance.

\subsection{Degree Distribution}
The next parameter studied is the degree distribution of all the networks. In a directed graph there are two types of degrees for a vertex: in-degree and out-degree . In-degree is the number of the edges that come into the vertex and out-degree is the number of edges that leave the vertex. In this study, the degree of a vertex is defined as the sum of its in-degree and out-degree. \\

The degree distribution is the probability of finding a vertex with the given degree. It signifies the probability of occurrence of junctions with 2, 4, 6, etc. number of roads intersecting to form that junction. From Figures \ref{fig:deg_distribution_1} and \ref{fig:deg_distribution_2} below, we observe that, as expected, the Erdos-Renyi model graphs exhibit exponential degree distribution as represented by the blue bell curve, but the same is not true for the network of a province and its corresponding network in the configuration model. However, the latter two networks exhibit the same degree distribution. This can be attributed to the fact that the configuration model network graph is generated using the in-degree and out-degree sequence of that province. 
All the provinces exhibit a similar-looking degree distribution. This can be attributed to the fact that we are using junctions as nodes and the roads forming the junctions as edges.

\begin{figure}[!htbp]
    \includegraphics[width=1\textwidth]{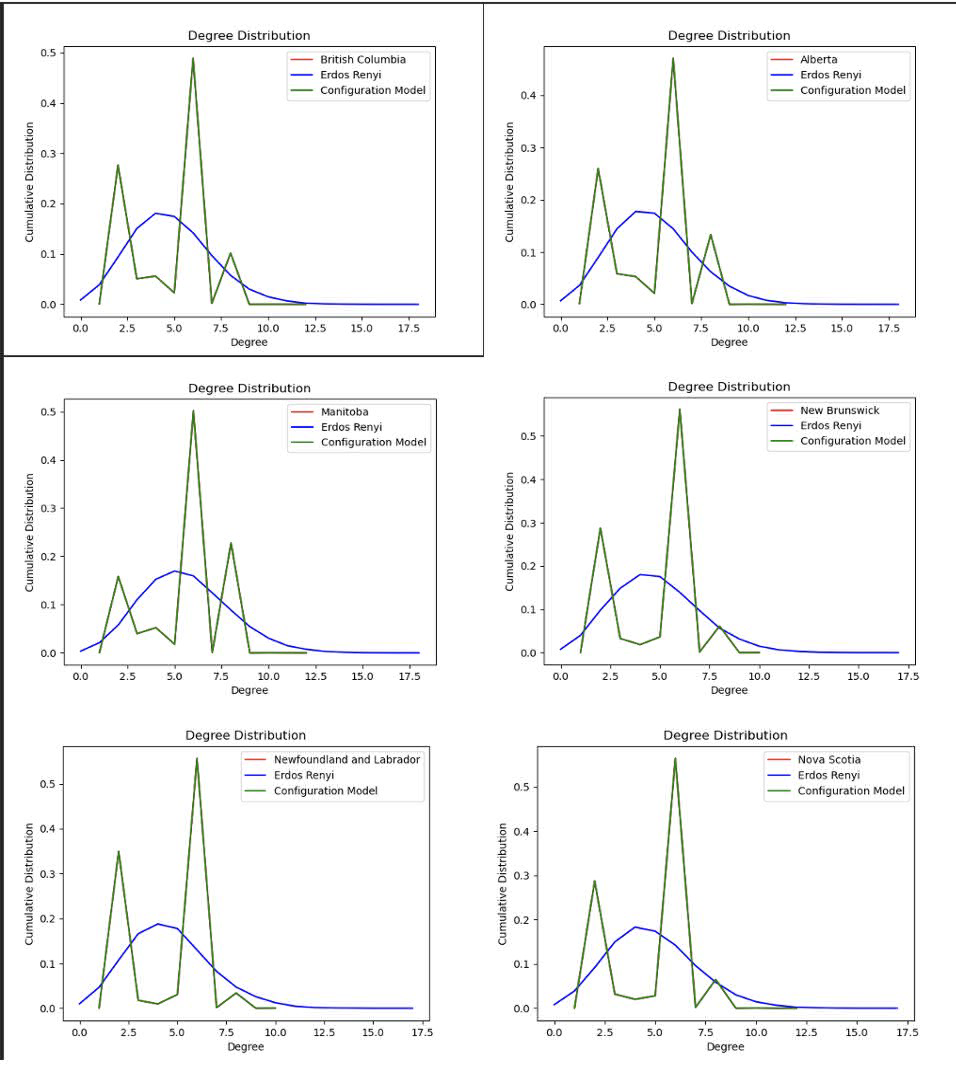}
    \caption{Degree distribution of each province}
    \label{fig:deg_distribution_1}
\end{figure}

\begin{figure}[!htbp]
     \includegraphics[width=1\textwidth]{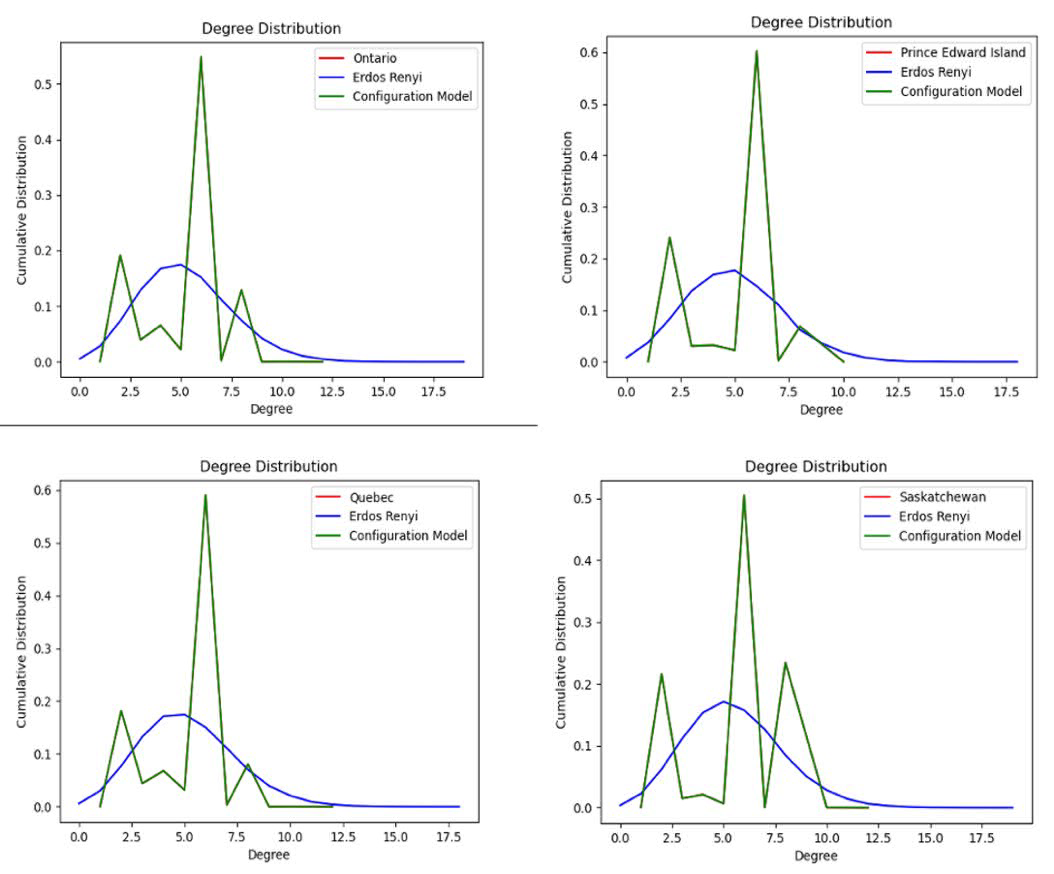}
    \caption{Degree distributions of each province}
    \label{fig:deg_distribution_2}
\end{figure}

\subsection{Average Degree of Neighbours}
The average degree of neighbours is a measure used to identify the kind of neighbours a node associates with. It is a measure to identify if there is interdependence from one intersection to another. One should expect for the actual network of the province, the average degree of neighbours increases as the degree of the node increases. This is because the purpose of the transport network is to support movement and to prevent congestion there is a low possibility that an intersection's neighbours have small degrees.\\

 From Figure \ref{fig:deg_neighbours} below, we observe that in all the province networks the average degree of nearest neighbours of a node increases with its degree, whereas for the random networks, it is relatively constant. This agrees with our expectation and the assortativity measure studied in the next section, lends support to this.\\
 
  Table \ref{table:shortest distance} shows the global average degree of neighbours for all the networks.

\begin{figure}[!htbp]
    \includegraphics[width=1\textwidth]{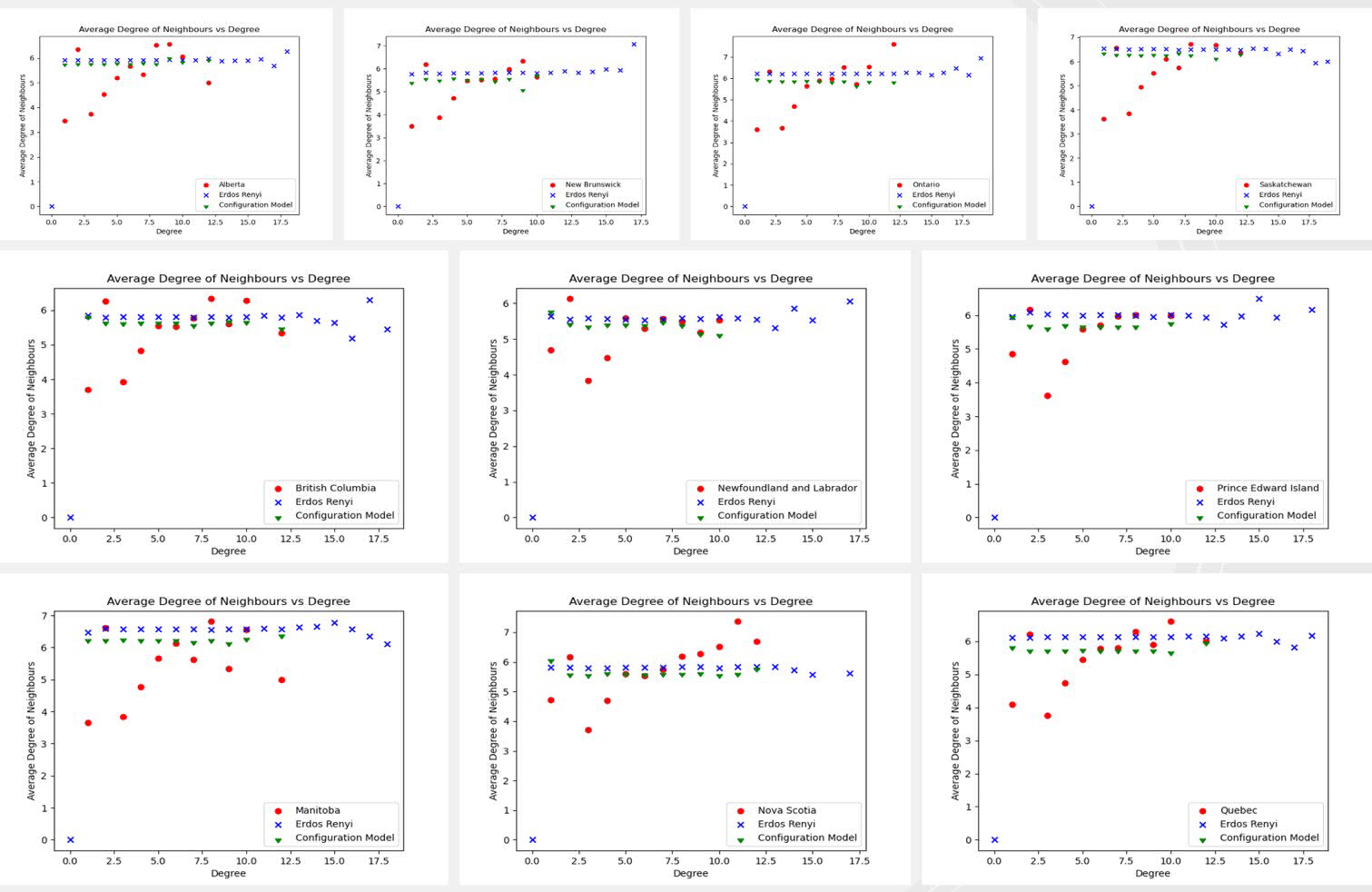}
    \caption{Average degree of neighbours}
    \label{fig:deg_neighbours}
\end{figure}

\begin{table}[!htbp]
    \label{table:Avg deg neighbours}
    \centering
    \begin{tabular}{lc}
        \toprule
        \textbf{Graph Name} & \textbf{Avg Degree of Neighbours} \\
        \midrule
        Nova Scotia & 5.76932931 \\
        Erdos\_Renyi\_Nova Scotia & 5.446416528 \\
        Configuration\_Model\_Nova Scotia & 5.684723332 \\
        Ontario & 5.648767403 \\
        Erdos\_Renyi\_Ontario & 5.959860155 \\
        Configuration\_Model\_Ontario & 5.889503094 \\
        Manitoba & 5.456051606 \\
        Erdos\_Renyi\_Manitoba & 6.208088796 \\
        Configuration\_Model\_Manitoba & 6.273989996 \\
        Quebec & 5.507781483 \\
        Erdos\_Renyi\_Quebec & 5.792333836 \\
        Configuration\_Model\_Quebec & 5.78562714 \\
        Alberta & 5.309459965 \\
        Erdos\_Renyi\_Alberta & 5.616876808 \\
        Configuration\_Model\_Alberta & 5.846647238 \\
        Saskatchewan & 5.60740285 \\
        Erdos\_Renyi\_Saskatchewan & 6.117146874 \\
        Configuration\_Model\_Saskatchewan & 6.310307503 \\
        New Brunswick & 5.274978991 \\
        Erdos\_Renyi\_New Brunswick & 5.580254002 \\
        Configuration\_Model\_New Brunswick & 5.528243765 \\
        British Columbia & 5.373841507 \\
        Erdos\_Renyi\_British Columbia & 5.46544186 \\
        Configuration\_Model\_British Columbia & 5.670169289 \\
        Prince Edward Island & 5.394721234 \\
        Erdos\_Renyi\_Prince Edward Island & 5.685891107 \\
        Configuration\_Model\_Prince Edward Island & 5.751521122 \\
        Newfoundland and Labrador & 5.18232645 \\
        Erdos\_Renyi\_Newfoundland and Labrador & 5.275580474 \\
        Configuration\_Model\_Newfoundland and Labrador & 5.422694977 \\
        \bottomrule
    \end{tabular}
    \caption{Average degree of neighbors by graph type and province}
\end{table}

\newpage
\subsection{Topological Assortativity}
In the previous section we observed that there is a positive correlation between degree of a node and average degree of its neighbours. Another parameter that estimates this correlation is Assortativity. Assortativity measures to which kind of neighbours a node prefers to attach. In other words it tries to answer the question on whether high degree nodes prefer to attach to other high degree node or not \cite{newman2003structure}. Following \cite{foster2010}, we have computed 
four different types of assortativity, using the correlation formula below:

\begin{equation*}
r(\alpha, \beta) = \frac{E^{-1} \Sigma_{i} (j_i^{\alpha} - \overline{j^{\alpha}})(k_i^{\beta} - \overline{k^{\beta}})}{\sigma^{\alpha}\sigma^{\beta}},
\end{equation*}

where $E$ is the set of all the edges in the network and $i$ is label of the $i$-th edge; the variables $\alpha$ and $\beta$ are in the set $\{in, out\}$, 
$j$ and $k$ are, respectively, the labels of the source and target vertex, $\overline{j^{\alpha}}$ is the mean of the $j_i^{\alpha}$ values taken over all the edges $E$, while $\sigma^{\alpha} = \sqrt{E^{-1} \Sigma_i(j_i^{\alpha} - \overline{j^{\alpha}})^2}$. The terms $\overline{j^{\beta}}$ and $\sigma^{\beta}$ have analogous definitons.

\begin{table}[!htbp]
    \centering
    \begin{tabular}{|p{0.5\textwidth}|p{0.125\textwidth}|p{0.125\textwidth}|p{0.125\textwidth}|p{0.125\textwidth}|}
        \toprule
        \textbf{Road Network ID} & \textbf{In-In} & \textbf{Out-Out} & \textbf{In-Out} & \textbf{Out-In} \\
        \midrule
        Nova Scotia & -0.0009 & -0.0009 & 0.0239 & 0.0004 \\
        Erdos\_Renyi\_Nova Scotia & -0.0001 & 0.0014 & 0.0044 & 0.0011 \\
        Configuration\_Model\_Nova Scotia & 0.0012 & 0.0007 & 0.0011 & 0.0009 \\
        Ontario & 0.1581 & 0.1587 & 0.1862 & 0.1572 \\
        Erdos\_Renyi\_Ontario & -0.0006 & 0.0013 & 0.0012 & 0.0001 \\
        Configuration\_Model\_Ontario & -0.0010 & -0.0009 & -0.0010 & -0.0008 \\
        Manitoba & 0.2119 & 0.2121 & 0.2366 & 0.2084 \\
        Erdos\_Renyi\_Manitoba & 0.0022 & -0.0017 & 0.0005 & -0.0001 \\
        Configuration\_Model\_Manitoba & 0.0011 & .0015 & 0.0014 & 0.0009 \\
        Quebec & 0.1197 & 0.1204 & 0.1591 & 0.1218 \\
        Erdos\_Renyi\_Quebec & -0.0007 & 0.0021 & 0.0008 & 0.0015 \\
        Configuration\_Model\_Quebec & 0.0007 & 0.0002 & 0.0008 & 0.0001 \\
        Alberta & 0.0.1268 & 0.0.1309 & 0.1590 & 0.1254 \\
        Erdos\_Renyi\_Alberta & 0.0003 & -0.0008 &  -0.0011 & 0.0009 \\
        Configuration\_Model\_Alberta &  -0.0004 & -0.0007 & -0.0007 & -0.0004 \\
        Saskatchewan & 0.1102 & 0.1110 & 0.1208 & 0.1087 \\
        Erdos\_Renyi\_Saskatchewan & -0.0006 & -0.0013 & -0.0035 & -0.0018 \\
        Configuration\_Model\_Saskatchewan & -0.0009 & -0.0010 & -0.0009 & -0.0010 \\
        New Brunswick & -0.0364 & -0.0372 & -0.0070 & -0.0316 \\
        Erdos\_Renyi\_New Brunswick & 0.0003 & 0.0013 & 0.0000 & 0.0057 \\
        Configuration\_Model\_New Brunswick & 0.0038 & 0.0032 & 0.0036 & 0.0034 \\
        British Columbia & 0.0603 & 0.0604 & 0.0847 & 0.0608 \\
        Erdos\_Renyi\_British Columbia & 0.0008 & -0.0017 & 0.0022 & -0.0021 \\
        Configuration\_Model\_British Columbia & 0.0008 & 0.0008 & 0.0009 & 0.0007 \\
        Prince Edward Island & -0.0027 & -0.0016 & -0.0027 & -0.0015 \\
        Erdos\_Renyi\_Prince Edward Island & 0.0003 & -0.0086 & 0.0004 & -0.0025 \\
        Configuration\_Model\_Prince Edward Island & 0.0216 & 0.0217 & 0.0486 & 0.0227 \\
        Newfoundland and Labrador & -0.1416 & -0.1429 & -0.1232 & -0.1382 \\
        Erdos\_Renyi\_Newfoundland and Labrador & -0.0025 & -0.0016 & 0.0050 & -0.0018 \\
        Configuration\_Model\_Newfoundland and Labrador & -0.0091 & -0.0091 & -0.0094 & -0.0088 \\
        \bottomrule
    \end{tabular}
    \caption{Assortativity Types}
    \label{table:assortativity}
\end{table}

From the table \ref{table:assortativity}, we observe that the both in-assortativity and out-assortativity coefficients are relatively high and are positive for all province networks barring the road network of Newfoundland and Labrador. This implies that there is assortative mixing in these networks except for the province of 
Newfoundland and Labrador and is consistent with the results from the graphs shown above.

\section{Network Resilience}
In the context of complex networks, network resilience involves studying how various types of networks, such as social networks, transportation systems, power grids, or computer networks, respond and adapt to disruptions (targeted attacks) or failures. Simulating attacks on a network can also be used to measure a network's resilience.\\

 There are two types network attack possible: one is a vertex-based attack and the other is an edge-based attack. Holme et al. \cite{holme2002attack} compared both types of attack for two real-world networks and four model networks, one of which is the Erdos Renyi model. Here we conduct a similar kind of study with an actual network and a corresponding random model network, generated using the configuration model. \\
 
 The description of the various kinds of attacks and the inferences we can draw from these simulations are addressed in the following sections. For all the computations we have assumed that the road networks are all undirected. 

\subsection{Vertex Based Attack}
A vertex-based attack identifies how the network behaves when vertices are removed  from a network. When we remove a vertex $v$, all the incident edges are also removed.\\

 There are many strategies available to remove vertices from a network. In this study, we discuss two of them: random removal of vertices and removal of vertices in descending order of vertex betweenness. The vertex betweenness measure identifies important vertices in the network. \\
 
 The  betweeeness centrality, $C_B(v)$, of a vertex $v$ is defined as  below (see \cite{brandes2001faster} for more details):\\
\begin{equation}
    C_B(v) = {\sum_{s \neq v \neq t} \frac{\sigma_{st}(v)}{\sigma_{st}}}\\
\end{equation}
\text{where:}
\begin{itemize}
 \item $\sigma_{st}$ is the total number of shortest paths from node $s$ to node $t$.
 \item $\sigma_{st}(v)$ is the number of those paths that pass through the node $v$.
 \item The sum is taken over all pairs of nodes $s$ and $t$, where $s \neq t$ and $s, t \neq v$.
\end{itemize}
 
 In random vertex removal, we repeatedly remove 10\% of the current vertices, selected at random from the network and to quantify network behaviour, the average shortest distance is then re-calculated for the residual (90\%) vertices. \\
 
  Another strategy we have explored is to remove repeatedly the top 10\% of the current set of vertices in decreasing order of vertex betweenness and recalculate the average shortest distance of the residual (90\%) vertices. \\
  
  Both strategies are repeated for the random graph, generated using the configuration model. \\
  
\begin{figure}[!htbp]
    \includegraphics[width=1\textwidth]{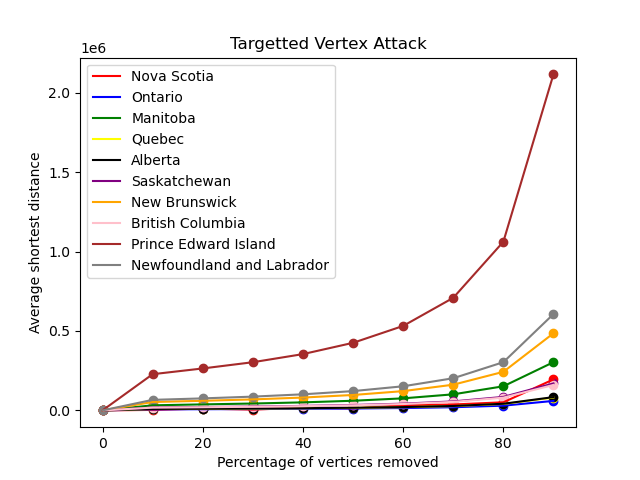}
    \caption{Average Shortest Distance vs Percentage of Nodes Removed}
    \label{fig:random_vertex_attack}
\end{figure}

  We have plotted a graph that shows how the average shortest distance varies for each province in the targeted removal of vertices (\ref{fig:random_vertex_attack}). From this figure, we can see that for Prince Edward Island there is a steep increase in the average shortest distance as vertices are removed. For the other provinces, the increase is gradual, implying other provinces are more resilient towards targeted 
  vertex-based attack. Another insight we can gather from this plot is that bigger networks are more resilient. Understandably so, as in the bigger networks there are multiple ways to get to a target node.\\

\begin{figure}[!htbp]
    \includegraphics[width=1\textwidth]{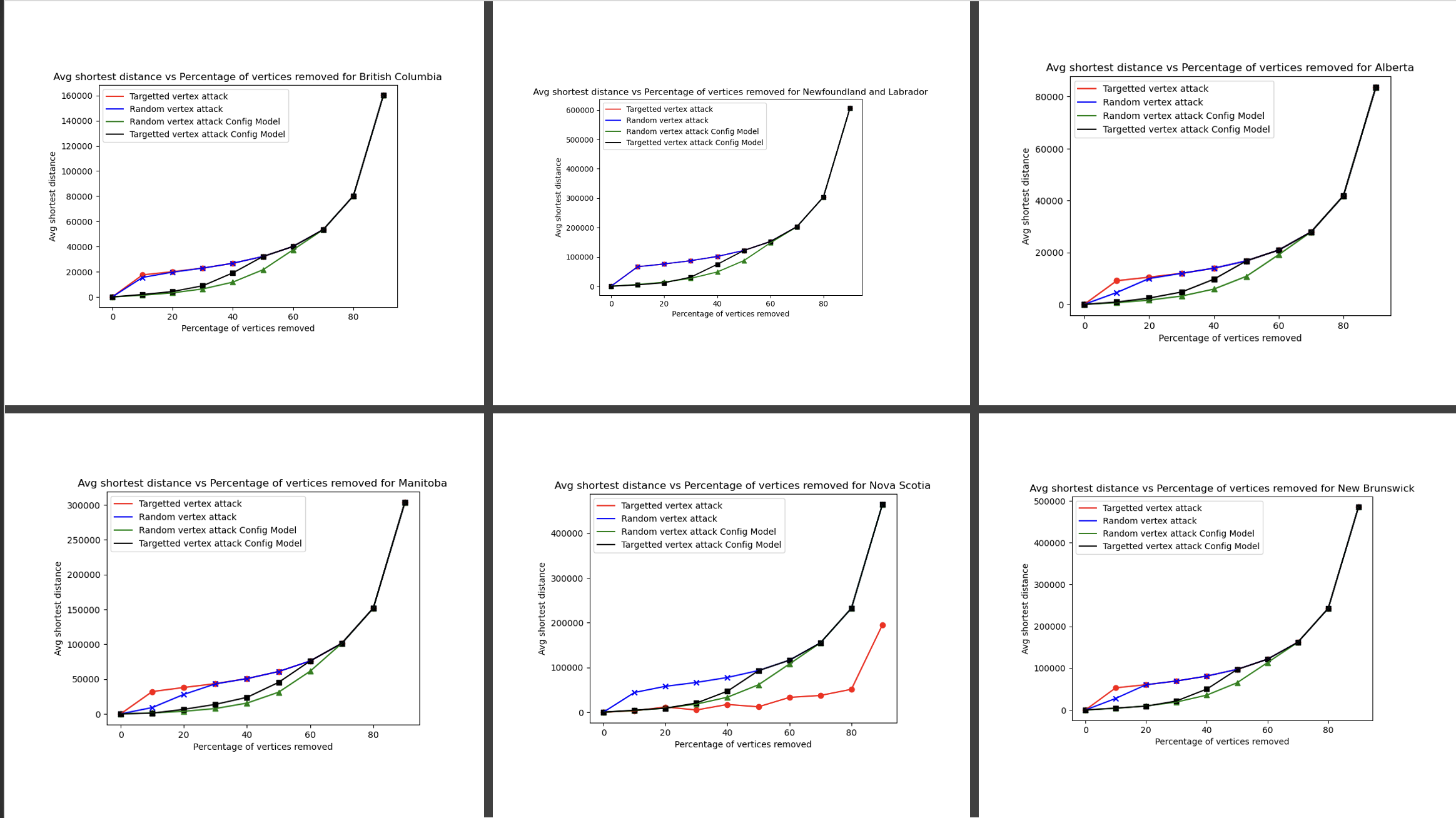}
    \caption{Random vs Targeted Vertex Attack}
    \label{fig:random_vs_targeted_v_attack_1}
\end{figure}

\begin{figure}[!htbp]
    \includegraphics[width=1\textwidth]{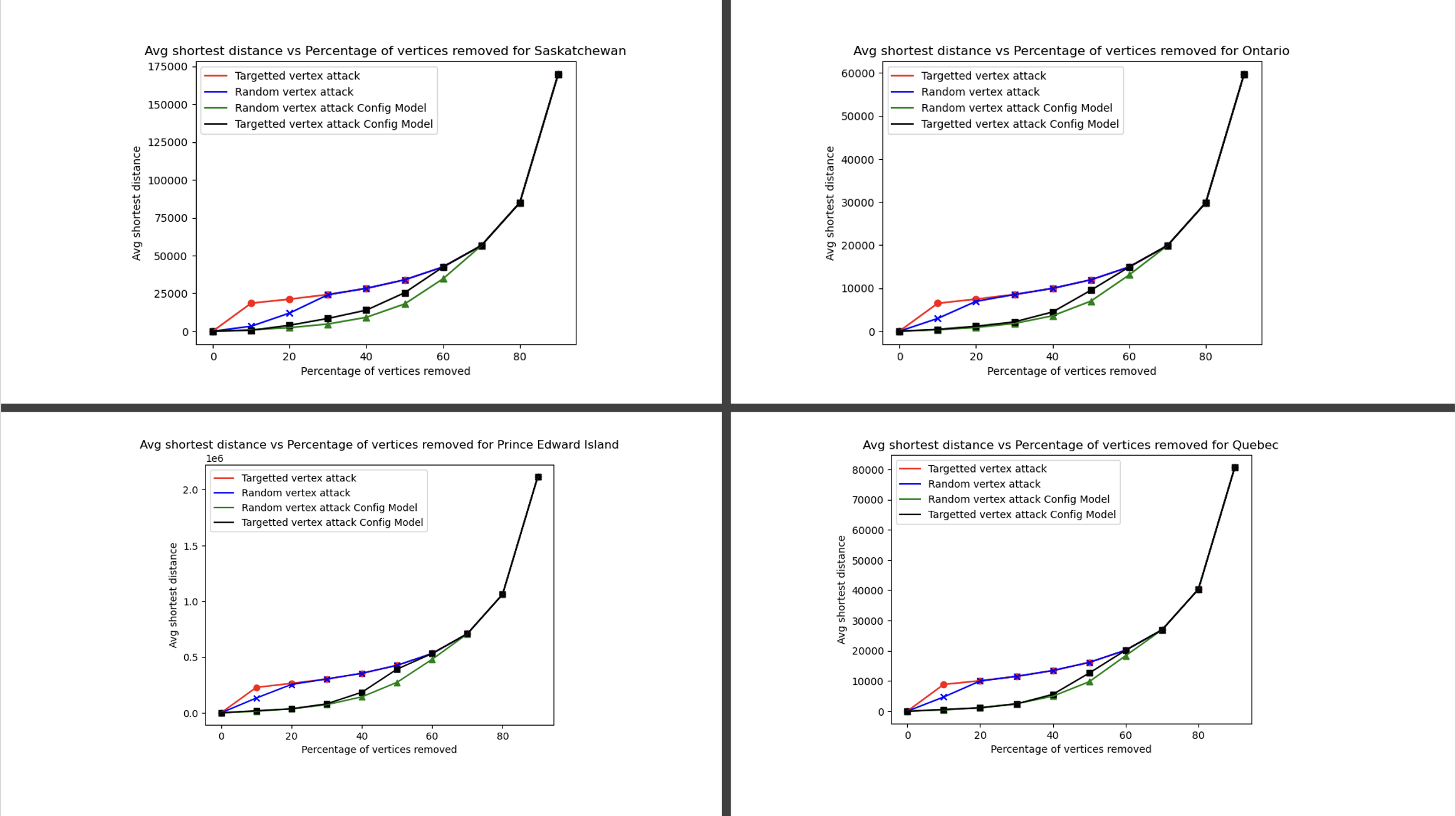}
    \caption{Random vs Targeted Vertex Attack}
    \label{fig:random_vs_targeted_v_attack_2}
\end{figure}

Figures \ref{fig:random_vs_targeted_v_attack_1} and \ref{fig:random_vs_targeted_v_attack_2} compare the impact on the average shortest distance of the random removal of vertices and targeted removal of vertices for the province networks and the random model networks.\\

 Given the definition of vertex betweenness, one should expect the average shortest distance to increase at first because important vertices are removed and then follow the trend of random removal as more and more vertices are removed. This behaviour is seen in almost all provinces except for Nova Scotia.\\
 
  Another observation is that in the random model, the impact of vertex removal is much more gradual when compared with the actual network. One possible explanation for this could be that in the actual network, the intersection and the road segments forming the intersection are interdependent. In contrast, in the random model, it is completely probabilistic.

\subsection{Edge Based Attack}
An edge-based attack quantifies how the network behaves on removal of edges from the network. Since the removal of an edge does not affect the vertices, repeated removal of edges leaves the network with isolated vertices. \\

As in the case of vertex removals, two strategies for removing edges are explored here:
one is the random removal of edges, where in every iteration 10\% of edges are selected at random and removed and is repeated until 90\% of the edges are removed. In every iteration the new average shortest distance between vertices of the network is calculated and the graph of how it varies is plotted. Second is the targeted removal of edges where again 10\% of the edges are removed in each iteration in the decreasing order of edge betweeness. Edge betweenness, $C_B(e)$, is calculated using the following formula \cite{brandes2001faster}:\\
\begin{equation}
C_B(e) = \sum_{s \neq t} \frac{\sigma_{st}(e)}{\sigma_{st}}
\end{equation}
\text{where:}
\begin{itemize}
    \item $C_B(e)$ represents the betweenness centrality of an edge $e$.
    \item $\sigma_{st}$ is the total number of shortest paths from node $s$ to node $t$.
    \item $\sigma_{st}(e)$ is the number of those paths that pass through the edge $e$.
    \item The summation is taken over all pairs of nodes $s$ and $t$ in the graph, where $s \neq t$.
\end{itemize}

\begin{figure}[!htbp]
    \includegraphics[width=1\textwidth]{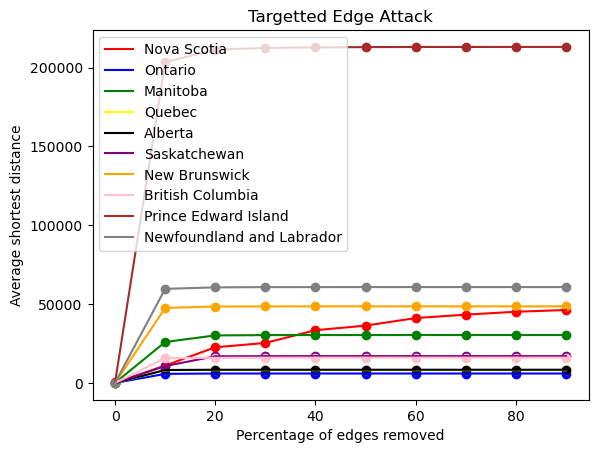}
    \caption{Avg Shortest Distance vs Percentage of edges Removed}
    \label{fig:random_edge_attack}
\end{figure}

\begin{figure}[!htbp]
    \includegraphics[width=1\textwidth]{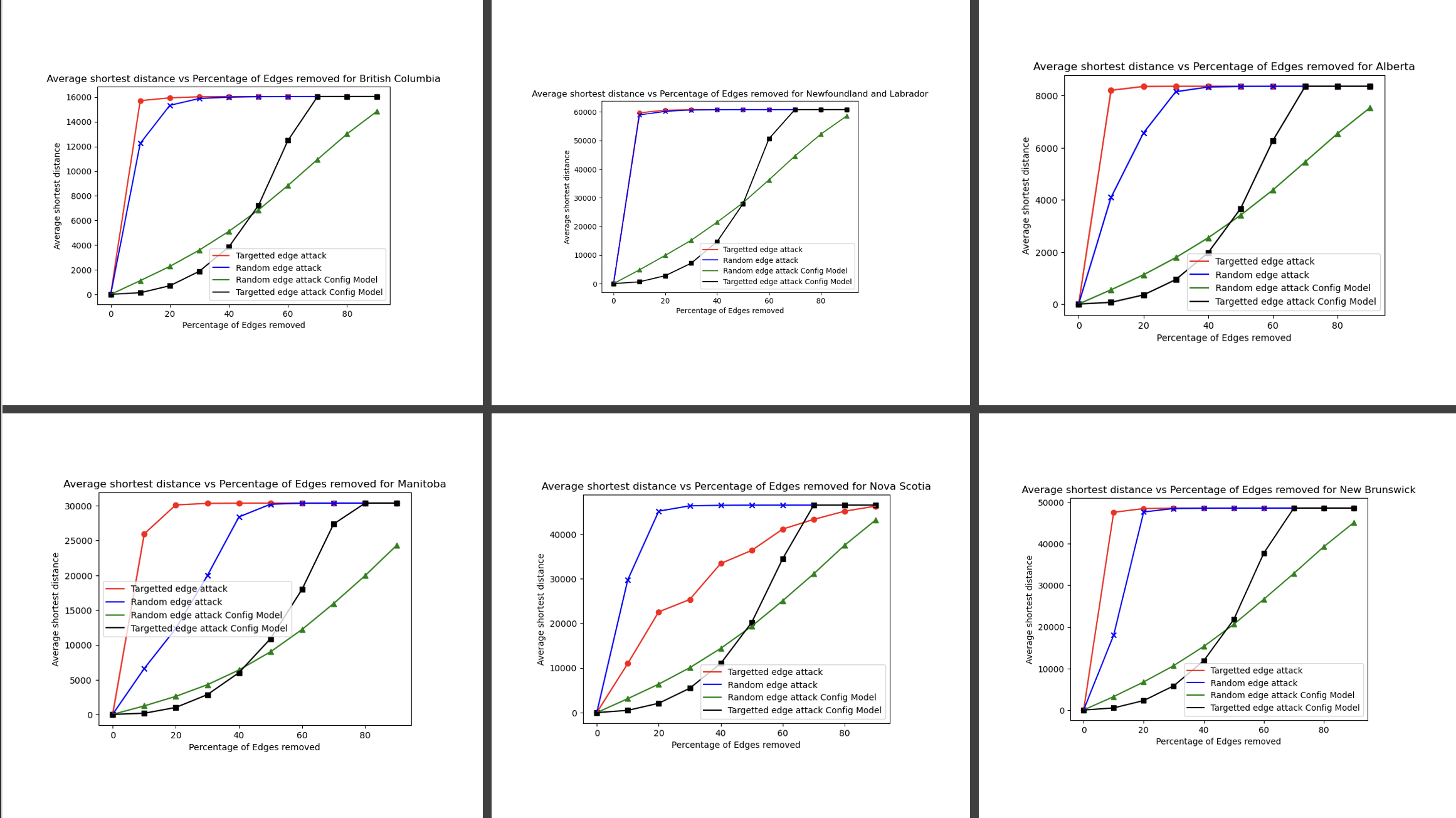}
    \caption{Random vs Targeted Edge Attack}
    \label{fig:random_vs_targeted_e_attack_1}
\end{figure}

\begin{figure}[!htbp]
     \includegraphics[width=1\textwidth]{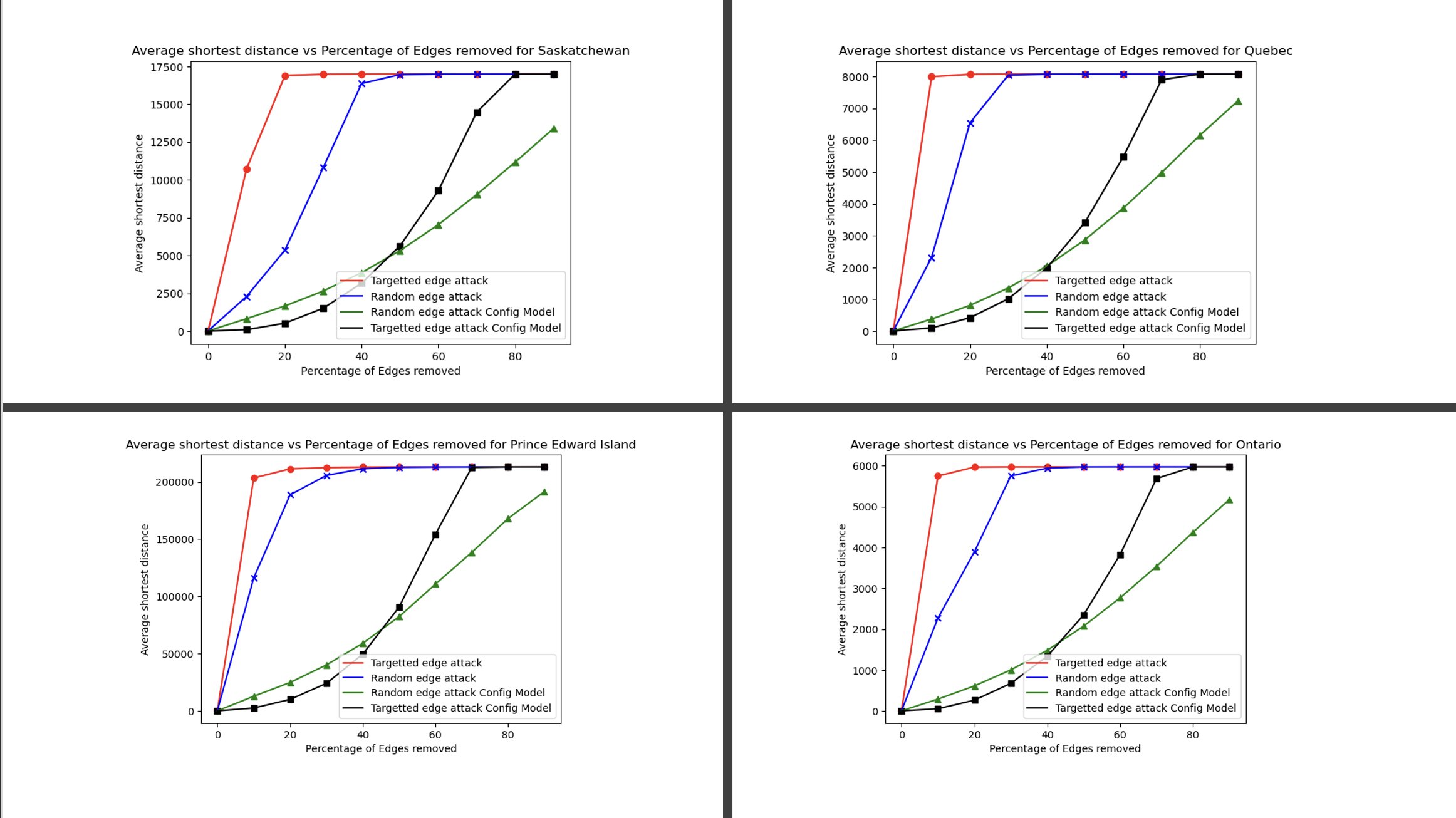}
    \caption{Random vs Targeted Edge Attack}
    \label{fig:random_vs_targeted_e_attack_2}
\end{figure}

Edge attacks as a whole have the same effect on province and random networks as vertex attacks seen in the previous section. Figure \ref{fig:random_edge_attack} shows that bigger networks are less susceptible to this kind of attack. Figures \ref{fig:random_vs_targeted_e_attack_1} and \ref{fig:random_vs_targeted_e_attack_2} also show that the average shortest distance increases gradually as compared with their respective actual province network. This strongly indicates that there is an interdependence between the vertices and edges of the network.

\section{Conclusion}
In this paper, the province networks of 10 provinces are extracted. Two random graphs, one using the Erdos Renyi model and the other using the configuration model with the same number of nodes and the same number of edges as the province network are generated. Parameters like clustering coefficient, degree distribution, average degree of neighbours and assortativity are compared among provinces and with their respective random graphs. An approach to quantify network resilience is discussed and is compared with one of the random models and among provinces. We found that the province network did not exhibit small-world properties, the larger the size of the province network more resilient it was to both vertex-based attacks and edge-based attacks and finally as expected there is an interdependence between edges and nodes of a network. We also noted that the configuration model was able to model the network more closely than Erdos-Renyi model.

\newpage
\bibliographystyle{plain}
\bibliography{referencesRoadNetwork}

\end{document}